\documentstyle[psfig,conf_iap,10pt]{article}
\begin{document}
\heading{%
%
Pre-DEIMOS Pilot Surveys for DEEP
%
} 
\par\medskip\noindent
\author{%
David C. Koo
}
\address{%
UCO/Lick Observatory, University of California, \\
Santa Cruz, CA 95064, USA
}

\begin{abstract}

DEEP is a multi-institutional program designed to undertake a major
spectroscopic survey of 10,000$^+$ field galaxies to $I \sim 23$ with
a new instrument (DEIMOS) on the Keck II 10-m telescope.  The
scientific goals include exploring galaxy formation and evolution,
mapping the large scale structure at moderate to high redshifts, and
constraining the nature and distribution of dark matter and cosmology.
DEEP is distinguished by securing spectra of sufficient quality and
resolution to extract rotation curves, velocity dispersions,
age estimates, and chemical abundances for a brighter subset of
galaxies.  While waiting for DEIMOS to be operational in 1999, the
first phase of DEEP science programs has concentrated on
LRIS observations of fields  observed with HST.
Recent highlights include redshift and kinematic studies of compact
galaxies, high redshift ($z \sim 3$) galaxies, and distant spirals.
\end{abstract}
\section{Introduction}
The 21st century promises many faint redshift surveys with the suite
of new 8-10 m class ground-based optical telescopes.  Besides
complementing data from space and other wavebands with critical
redshifts, the high S/N and high spectral-resolution from 8-10 m class
telescopes provide three new, powerful diagnostics for the analysis of
distant galaxies, namely internal velocities (and hence masses when
size is used), abundances, and age estimates.  These parameters are
clearly important, independent probes of galaxies in the early
universe with solid links to theoretical simulations. Finally, because
both galaxy evolution and their large scale patterns are very likely
to be complex, surveys with large samples will be needed to address
these problems.

\section{What is DEEP?}

To meet the challenge, 
the Deep Extragalactic Evolutionary Probe
(DEEP: more details on participants and programs at URL: {\bf
http://www.ucolick.org/ $\sim$deep/home.html})
was initiated over 6 years ago
as  a project designed to gather spectral
data for over 10,000 faint field galaxies \cite{Mou93},\cite{Koo95a},
\cite{Koo98} 
using the Keck II 10-m
telescope and a new spectrograph for Keck II (DEIMOS: DEep Imaging
Multi-Object Spectrograph; more information is provided at 
URL: {\bf
http://www.ucolick.org/ $\sim$loen/Deimos/deimos.html} and
contribution
from Davis in these proceedings).

A distinguishing aspect of DEEP is that the survey aims to gather
not only very faint redshifts, but also internal kinematic data in the form of rotation curves or
line widths, as well as line strengths  sensitive to star formation
rates, gas conditions, age,  and metallicity.

\section{Highlights of First Phase DEEP Projects}

While waiting for the completion of DEIMOS so that the major DEEP
survey of 10,000$^+$ galaxies can begin (see Davis contribution), we
have been undertaking a number of smaller, pilot-style projects with
the existing Low Resolution Imaging Spectrograph (LRIS: \cite{Oke95})
to determine what is feasible with Keck and thus to help refine the
scope of the main DEEP survey. To maximize the scientific returns for
our relatively small samples (currently over 500 galaxies), we
observed fields where HST WFPC2 images already exist, including the
HDF and flanking fields \cite{Low97}, \cite{Phi97}, \cite{Guz97},
\cite{Vogt97}; the Groth Survey Strip \cite{Koo96}, \cite{Vogt96}, and
Selected Area 68. Such HST images provide morphology data and also the
structure, size, and inclination data needed to convert kinematic
observations from Keck into direct measures of mass.  Our data is
still largely being reduced, but we have already achieved a redshift completeness of
97\% for a 200 galaxy sample reaching a limit of $I \sim 23.5$.
Overall, our findings have reassured us that our major DEIMOS survey
is not only feasible, but that kinematics will indeed be a powerful
additional dimension of study. Our work with line strengths is not yet
mature enough to be presented, so the following will focus on
highlights of our kinematic surveys.

\begin{figure}
\centerline{\vbox{
\psfig{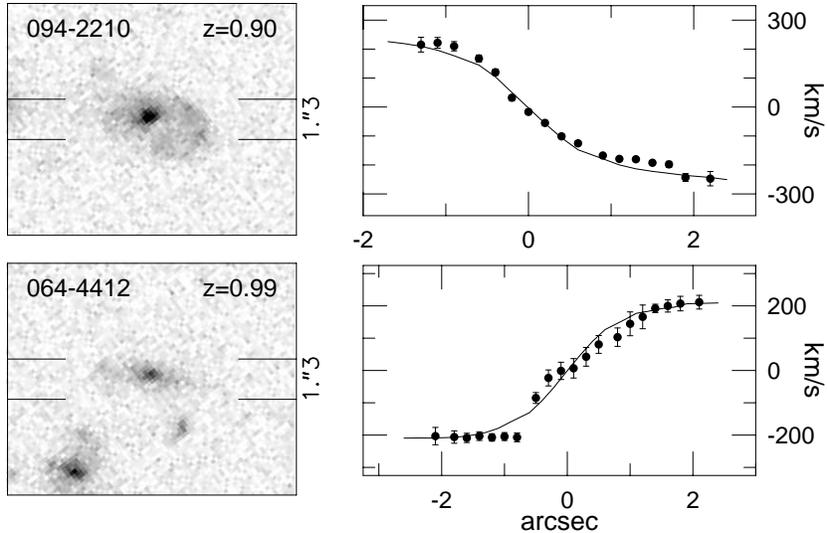}
}}
\caption[]{
Examples of the rotation curves measured for 
two high redshift galaxies, the upper with total $I \sim 21.4$ and the lower
with $I \sim 22.4$ \cite{Vogt97}.
}
\end{figure}

\subsection{Rotation Curves of Distant Spirals}

As seen in Fig. 1, we have clearly demonstrated that emission-line
rotation curves of likely spirals can be observed to redshifts near $z
\sim 1$ for galaxies as faint as $I \sim 22$ with one to two hour
exposures \cite{Vogt96}.  Based on 16 galaxies so far, we find little
evidence for any major change ($<0.6$ mag) in the zero-point of the
optical Tully-Fisher relation \cite{Vogt97}.  These results are in
stark contrast to claims for more extensive evolution of 1.5 mag
to 2.0 mag \cite{Rix97}, \cite{Sim98} for very blue galaxies. Larger
samples will be needed to understand the causes (e.g., luminosity or
color) of these differences.

\subsection{Emission Line widths}

Though rotation curves are preferable, the vast majority of very faint
galaxies are too small to yield more than line widths as kinematic
data.  Except for very bright galaxies that might yield absorption
line widths, emission lines are used.  Though winds, dust obscuration,
and poor representation of the gravitational potential by the luminous
star formation regions may all invalidate the use of the emission line
widths for probing the gravitational potential, the available data for
compact HII galaxies, which are the most likely to suffer from these
problems, nevertheless show a fairly tight correlation with a ratio of
0.7 +/- 0.1 between such optical line-width measures and radio
measures of HI motions \cite{Tell93} that should be sampling well the
total mass of such systems.  

Assuming line widths, after an upward correction of 40\% \cite{Rix97}, are generally
meaningful measures of the true gravitational potential, and adding
HST sizes, we are able to obtain masses.  At least for blue compact
galaxies, we then find that luminosity alone can be a very poor gauge
of their masses, i.e., the M/L ratio can vary enormously \cite{Guz96},
\cite{Phi97}, \cite{Guz97}.  For some, we even needed the the High
Resolution Echelle Spectrograph (HIRES: see \cite{Vogt94}) to resolve
velocity widths smaller than 30 km-s$^{-1}$ \cite{Koo95b}.
Fortunately for the DEEP program, the vast majority of faint galaxies
have line widths that should be resolved at with DEIMOS.  The key point
is that the lack of correlation between optical luminosity and mass,
i.e. stable M/L, demonstrates the necessity, usefulness, and promise
of kinematics as an important new dimension to discern the evolution
of different galaxy populations.

\subsection{Very High Redshift $z \sim 3$ Galaxies}

A major advance with Keck has been the dramatic
demonstration  that very high redshift ($z
\sim 3$) galaxies chosen with colors
can be confirmed spectroscopically \cite{Stei96a}. The DEEP team
has extended the pioneering efforts \cite{Stei96b} in the
Hubble Deep Field (HDF) by pushing over one magnitude fainter,
using redder ``dropouts'' to reach higher redshifts and higher
levels of completeness, and adopting higher spectral resolutions to
improve kinematic measurements \cite{Low97}.  Based on the
evidence so far \cite{Low98}, \cite{Pet98}, 
the high redshift galaxies may also
be small mass systems that become dwarfs today or that later merge to
form more massive galaxies \cite{Low97}, \cite{Som98} 
instead of being only the cores of massive
galaxies. 

\section{Summary}

Our various first phase, pilot programs with LRIS clearly show 1) the need and power
of kinematics for galaxy surveys, 
2) the feasibility of reaching $I \sim 23$ or fainter
with respectable completeness in one to two hour exposures with Keck,
and 3) the
accessibility of the very high redshift universe.  Thus the
definition, feasibility, and value 
of the second phase DEIMOS DEEP program is no longer merely 
in the speculative proposal stage.

\acknowledgements{ DEEP was initiated by the Berkeley Center for
Particle Astrophysics (CfPA), and has been supported by various NSF,
NASA, and STScI grants. The senior members of DEEP have managed the
project, but I would like to give special thanks to our talented pool
of more junior astronomers over the years (see DEEP URL for names),
without whom the results presented here would not have been possible.
}



%
\vfill
\end{document}